\documentclass[10pt,twocolumn,aps,pre,superscriptaddress]{revtex4-1} %

\usepackage{graphicx}
\usepackage{dcolumn}
\usepackage{amsmath}
\usepackage{amssymb}
\usepackage{bm}
\usepackage{color}
\usepackage{ulem}
\newcommand{\cref}[1]{(\ref{#1})}

\begin{document}

\title{The Raspberry Model for Hydrodynamic Interactions Revisited. II.\protect\\ The Effect of Confinement}

\author{Joost de Graaf}
\email{jgraaf@icp.uni-stuttgart.de}
\affiliation{Institute for Computational Physics (ICP), University of Stuttgart, Allmandring 3, 70569 Stuttgart, Germany}

\author{Toni Peter}
\affiliation{Institute for Computational Physics (ICP), University of Stuttgart, Allmandring 3, 70569 Stuttgart, Germany}

\author{Lukas P. Fischer}
\affiliation{Institute for Computational Physics (ICP), University of Stuttgart, Allmandring 3, 70569 Stuttgart, Germany}

\author{Christian Holm}
\affiliation{Institute for Computational Physics (ICP), University of Stuttgart, Allmandring 3, 70569 Stuttgart, Germany}

\date{\today}

\begin{abstract}
The so-called `raspberry' model refers to the hybrid lattice-Boltzmann (LB) and Langevin molecular dynamics scheme for simulating the dynamics of suspensions of colloidal particles, originally developed by [V. Lobaskin and B. D{\"u}nweg, New J. Phys. \textbf{6}, 54 (2004)], wherein discrete surface points are used to achieve fluid-particle coupling. In this paper, we present a follow up to our study of the effectiveness of the raspberry model in reproducing hydrodynamic interactions in the Stokes regime for spheres arranged in a simple-cubic crystal [L. Fischer, \textit{et al.}, J. Chem. Phys. \textbf{143}, 084107 (2015)]. Here, we consider the accuracy with which the raspberry model is able to reproduce such interactions for particles confined between two parallel plates. To this end, we compare our LB simulation results to established theoretical expressions and finite-element calculations. We show that there is a discrepancy between the translational and rotational mobility when only surface coupling points are used, as also found in Part I of our joint publication. We demonstrate that adding internal coupling points to the raspberry, can be used to correct said discrepancy in confining geometries as well. Finally, we show that the raspberry model accurately reproduces hydrodynamic interactions between a spherical colloid and planar walls up to roughly one LB lattice spacing.
\end{abstract}

\maketitle

\section{\label{sec:intro}Introduction}

\begin{figure}[!htb]
\begin{center}
\includegraphics[scale=0.75]{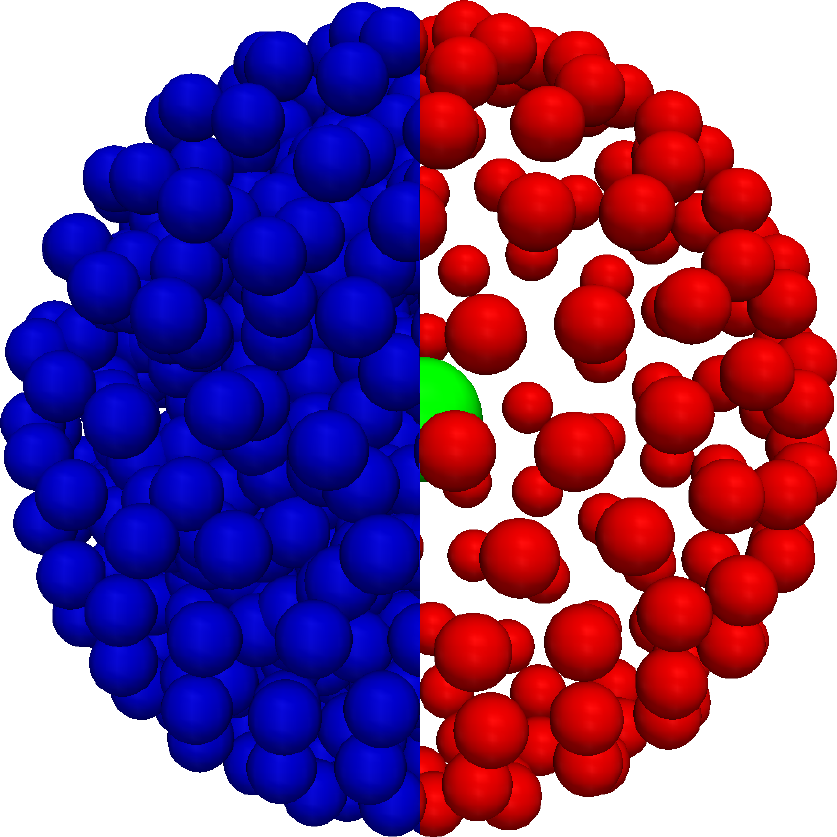}
\end{center}
\caption{\label{fig:rasp}(color online) Representation of the structure of the raspberry models used in our simulations, filled (left) and hollow (right), respectively. The central bead to which all other beads are connected \textit{via} rigid bonds is shown using a green sphere. The blue spheres represent the beads that form the filled raspberry and the red ones give the surface beads used for the hollow variant. The radius of the beads is chosen to be smaller than the typical effective hydrodynamic radius to help visualize the internal structure.}
\end{figure}

Here, we continue our investigation into the so-called `raspberry' model, which was first introduced by Lobaskin and D{\"u}nweg.~\cite{lobaskin04} The raspberry model allows for the incorporation of a colloidal particle into the lattice-Boltzmann (LB) algorithm by discretizing the surface into point particles and utilizing the coupling scheme of Ahlrichs and D{\"u}nweg~\cite{ahlrichs99} for these points. The model derives its name from the discretized nature of the colloid's surface, which resembles a raspberry, when represented by molecular-dynamics (MD) beads, see Fig.~\ref{fig:rasp} (right) for the original `hollow' variant~\cite{lobaskin04} and Fig.~\ref{fig:rasp} (left) for the improved `filled' raspberry introduced in Ref.~\cite{fischer15}. 

In Ref.~\cite{fischer15} we demonstrated the raspberry's ability to accurately capture hydrodynamic interactions in a simple-cubic arrangement. We discussed how the point coupling leads to an effective hydrodynamic radius that can be fitted for. As shown by Ollila~\textit{et al.}~\cite{Ollila12,ollila13} there is a mismatch between the effective hydrodynamic radii when the coupling points are placed on a spherical shell. Ollila~\textit{et al.} contend that this is related to the porosity of the raspberry particle, also see Part I~\cite{fischer15}, that is inherent to this coupling method. We showed in Ref.~\cite{fischer15} that one can effectively match the raspberry to results for solid particles within a reasonable parameter regime, via a procedure which we dubbed `filling + fitting.' That is, a raspberry particle with internal coupling points (filled), for which effective hydrodynamic radii are established via hydrodynamic experiments (fitting), shows excellent \textit{numerical} correspondence for the two fitted hydrodynamic radii. Moreover, the results agree well with theoretical predictions for a solid particle in the Stokes' limit, in the intermediate and long-time regime.

In this paper, we study the confining geometry of two parallel plates, to further demonstrate the quality of the `filling + fitting' procedure. This geometry is of particular interest, since to the best of our knowledge, no systematic study of the quality of the particle-coupling method of Ref.~\cite{lobaskin04} has been attempted in which the level of confinement was the crucial parameter. A two plate geometry has been applied in various works, see, for example, Refs.~\cite{ollila13,Mackay13b}, but always with the intention to extrapolate the results to infinite plate separations or extract bulk properties. For our study of small plate separations (strong confinement) in this manuscript, we use a variety of literature results to compare our results against. 

The first analytic expressions for the mobility of the sphere-plate system were derived by Faxen~\cite{faxen22,happel83}, who restricted his study to spheres constrained to the mid-plane. Brenner derived an expression for mobility of a sphere close to a single wall~\cite{brenner61} (in the direction perpendicular to the wall), which was later extended to a parallel-plate geometry in Refs.~\cite{lobry96,lin00}. These publications (\cite{faxen22,happel83,brenner61,lobry96,lin00}) are but the tip of the iceberg when it comes to the semi-analytic and numerical work that has been carried out on these and comparable geometries. Other works include: semi-analytic calculations,~\cite{oneill64,liron76,cichocki98,kalinay14} numerical simulations that exploit analytic approximations to capture hydrodynamic behavior,~\cite{dufresne01,swan11,pasol11,vonhansen11,bleibel14,mitchell14} and simulations that employ coarse-grained hydrodynamic solvers to fully resolve hydrodynamic interactions.~\cite{padding10,nikoubashman13,tatsumi13,dubov14} In addition, numerous experimental studies into the diffusive behavior of spheres under confinement,~\cite{volpe10,brettschneider11,dettmer14a,dettmer14b,lisicki14} as well as other shapes,~\cite{chakrabarty13,jones13,chakrabarty14} have been undertaken. However, it goes beyond the confines of this introduction to fully list and do justice to all these investigations.

In the following, we use some of the above literature results to prove that the hollow and filled raspberry reproduce Stokesian hydrodynamic interactions for a sphere confined between two parallel plates. As in Part I, the `filling + fitting' procedure is shown to resolve the issue of the mismatch between the effective hydrodynamic radii that the hollow variant of the raspberry suffers from, to within acceptable numerical tolerance. We also show that hydrodynamic interactions are properly captured by the raspberry-LB method up to one lattice spacing away from either wall. This is a reasonable result, since lubrication corrections for this point-coupling method have not yet been established. In the right limits, we were able to demonstrate that the mean-square displacement also yields the proper position dependence of the translational mobility in a thermalized LB fluid. Finally, we used our high-quality numerical data to establish empirical relations for this position dependence for cases where literature values were not available or did not extend over the full range of plate separations.

The remainder of this manuscript is structured as follows. In Section~\ref{sec:methods} we describe the aspects of our simulation methods that were not covered in Part I~\cite{fischer15} in detail. Section~\ref{sub:exper} describes the various hydrodynamic experiments that we performed to determine the properties of the raspberry model. For certain experiments we used finite-element calculations as the basis for comparison, the specifics of these are provided in Section~\ref{sub:FE}. We provide a summary of the notations used throughout the text in Section~\ref{sub:notation} to aid the reader when going through the manuscript. In Section~\ref{sec:result} we list our main results. These are discussed and related to previous studies and our work in Ref.~\cite{fischer15} in Section~\ref{sec:disc}. Finally, we give a summary, conclusions, and an outlook in Section~\ref{sec:conc}.

\section{\label{sec:methods}Methods}

In this section, we outline the modeling approaches used to determine the hydrodynamic properties of a colloid. We have split this into subsections detailing aspects of the hydrodynamic experiments performed to extract the mobility of the raspberry, the finite-element calculations, and a reference list of the input parameters and measured quantities. For details on the construction of the raspberry model and the Molecular Dynamics (MD) and lattice-Boltzmann (LB) parameters employed in our investigation, we refer the reader to Ref.~\cite{fischer15}.

\subsection{\label{sub:exper}Hydrodynamic Experiments}

\begin{figure}[!htb]
\begin{center}
\includegraphics[scale=1.0]{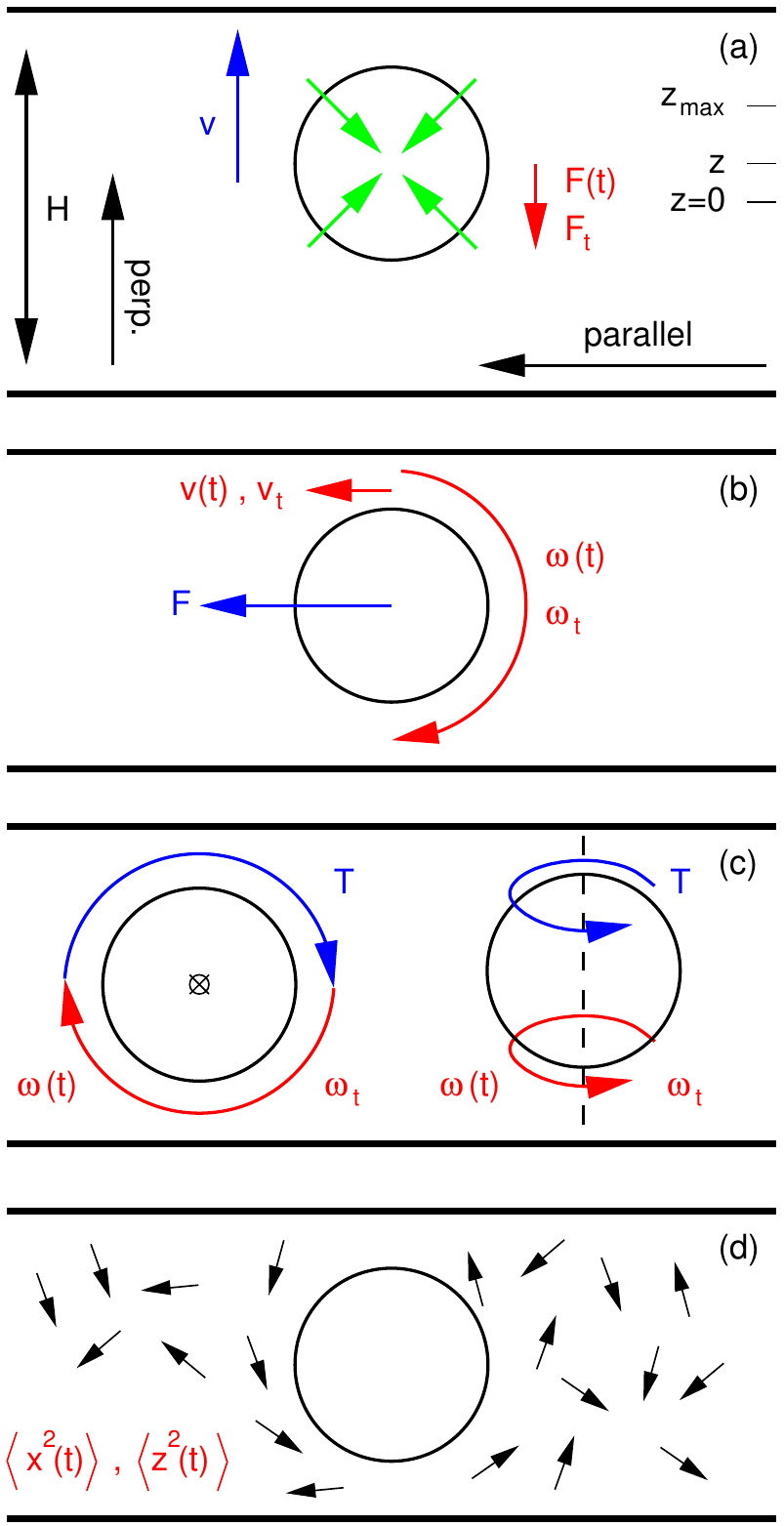}
\end{center}
\caption{\label{fig:plate}(color online) Visualization of the various hydrodynamic experiments carried out between two parallel plates, separated by a distance $H$. Here, a two-dimensional (2D) cross-section of the setups is shown. The blue arrows and symbols denote quantities applied to the fluid and raspberry, the red arrows and symbols indicate measured quantities, the green arrows indicate a spatially fixed raspberry. The black arrows signify a thermalized fluid. We refer to the text for a description of the experiments and applied and measured quantities.}
\end{figure}

To assess the quality of the raspberry approximation in modeling the hydrodynamic properties of a colloid we performed several experiments. In the following, we use the term `quiescent' to describe an un-thermalized (non-fluctuating, deterministic) LB fluid. We set up the system of a particle confined between two parallel plates as follows. A channel of height $H$ was simulated using $H + 2$ lattice points along the $z$-axis (one layer of boundary cells on either side). In the $xy$-plane the box had an extent $L$ in both directions and periodic boundary conditions were applied in these directions. We used finite-size scaling to remove any dependence on $L$, \textit{i.e.}, we varied the box size and fitted to the result for $L \uparrow \infty$; typically the results were sufficiently converged for $L > 5H$. The bounce-back boundary condition~\cite{frisch86} was applied to the first and final layer of points in $z$-direction, leading to an effective hydrodynamic channel height of $H$ with a no-slip surface. The height was confirmed by Hagen-Poiseuille flow measurements, by comparing to the theoretical fluid flow profile and fitting for the channel height. For all experiments the raspberry was initialized at a height $z$, measured with respect to the center of the channel ($z=0$), see Fig.~\ref{fig:plate}. 

The following experiments for a spherical particle confined between two parallel plates were performed, as detailed below. In Ref.~\cite{fischer15} we discuss in detail the Reynolds number ($Re$), P{\'e}clet number ($Pe$), Schmidt number ($Sc$), as well as other dimensionfree constants at which these experiments were carried out. For the purposes of this manuscript, it suffices to know that the typical values of these numbers are $Re < 0.1$, $Pe > 100$, and $Sc > 1000$.

\begin{itemize}

\item A \textit{velocity} experiment in a \underline{quiescent} fluid, see Fig.~\ref{fig:plate}(a). The particle was kept fixed at the initial position. Its surface velocity was held constant, by resetting the velocity in every time step. Thus an effective velocity boundary condition is obtained, without the particle changing position. The mobility was subsequently determined by measuring the force on the particle in the stationary state. This approach is similar to the one employed for the finite-element calcultions in Section~\ref{sub:FE}. Since the particle does not move, the position dependence of the translational mobility parallel and perpendicular to the wall could be accurately assessed.

\item A \textit{force} experiment in a \underline{quiescent} fluid, see Fig.~\ref{fig:plate}(b). The particle was allowed to move freely and a constant force was applied to its center of mass. The position dependent mobility for movement parallel to the walls could be determined by extracting the terminal velocity. This terminal velocity was averaged over at least 10 periods of the (slight) oscillation in this parameter due to interpolation artifacts. It proved unnecessary to fix the particle in the $z$ direction, since there was no appreciable vertical drift. The presence of the boundary breaks the symmetry of the system, when the particle is not in the center of the channel. Therefore, there can be rotational and translational cross-coupling terms in the hydrodynamic mobility tensor (HMT).~\cite{goldman67} To examine this we also measured the angular velocity of the particle.

\item A \textit{torque} experiment in a \underline{quiescent} fluid, see Fig.~\ref{fig:plate}(c). A constant torque was applied to the particle about an axis parallel (Fig.~\ref{fig:plate}(c), left) or perpendicular (Fig.~\ref{fig:plate}(c), right) to the walls and the angular velocity was measured. Again it proved unnecessary to fix the particle in the $z$ direction. For rotation about an axis parallel to the wall, we also fixed the particle's position, to examine the effect of cross coupling.

\item A mean-square displacement (MSD) experiment in a \underline{thermalized} fluid, see Fig.~\ref{fig:plate}(d). The system was equilibrated until the particle fluctuated with the proper imposed fluid temperature. Then over many integration cycles (typically $10^{8}$ per run) and for several runs, we kept track of the position of the particle. The positions were binned according to the following procedure. Each bin was made half a lattice spacing (0.5$\sigma$) in width for the direction of interest. All sub-trajectories of our runs that originated in a specific bin were assigned to that bin and followed until they exited the bin. To ensure a minimum smearing out of the position dependence, we selected only those parts of the trajectories that went through a slice 0.1$\sigma$ in diameter around the center of the bin. This eliminated the trajectories that moved in and out of the bin but never crossed the center. For each selected trajectory, we determined the MSD and averaged this over all the relevant trajectories. Each individual MSD was obtained by examining such a trajectory over 50,000 integration steps from its `starting' point. Because we used a low temperature ($k_{\mathrm{B}}T = 0.01\epsilon$), we were able to access the long-time diffusion regime without the particle traveling beyond the confines of the relevant bin, despite the long MSD sampling length, see Part I.~\cite{fischer15} This is necessary, as the Ahlrichs and D{\"u}nweg coupling method~\cite{ahlrichs99} used to thermalize the raspberry particles, does not accurately reproduce the short-time diffusion that is present for a system that satisfies Stokes' equations for all times, see the discussion in Ref.~\cite{fischer15}.

\end{itemize}

\subsection{\label{sub:FE}Finite-Elements Calculations}

In order to compare the quality of our results for the raspberry particles to reference curves, we utilized literature values from analytic and numerical studies whenever possible. However, for the sphere confined between two plates, literature results were often not available, to the best of our knowledge. Therefore, we performed finite-element calculations using the COMSOL 4.4 Multi-Physics Modeling Software to establish reference data.

We set up a system with a spherical colloid of radius $R = 0.5$ $\mu$m. The sphere was located on the symmetry axis of a cylindrical fluid domain of height $H$, bounded on top and bottom by two no-slip plates (the bases of the cylinder). The cylindrical domain was bounded on its side by open boundary conditions (zero normal stress) for the fluid. We solved Stokes' equations for an incompressible fluid with density $\rho = 10^{3}$ kg$\,$m$^{-3}$ and dynamic viscosity $\eta = 10^{-3}$ kg$\,$m$^{-1}\,$s$^{-1}$ on this domain. 

To determine the various mobilities we imposed the following boundary conditions on the sphere. We solved for the force and torque exerted by the fluid and plates on the sphere, to which we applied a constant (angular) velocity boundary condition, to obtain the translational and rotational mobility, respectively. To ensure that the laminar flow condition is satisfied, we used low values of the (angular) velocity of the sphere: $v = 10^{-6}$ m$\,$s$^{-1}$ and $\omega = 10^{-6}$ s$^{-1}$. Typically, we found Reynolds numbers $Re^{T}$ and $Re^{R} \ll 10^{-5}$.

We varied the position of the sphere with respect to the plates along the cylindrical domain's axis keeping the sphere's position fixed in the other directions to determine the $z$ dependence of the mobility. Similarly, we examined the plate-separation dependence $H$ of the mobility by keeping the sphere fixed in the center of the cylinder and moving the plates outward. The diameter of the cylinder was varied to remove finite-size effects, but we found that a diameter of $10H$ was sufficient to guarantee convergence.

Whenever possible, we exploited the rotational symmetry of the system to minimize the number of elements required, \textit{i.e.}, for perpendicular movement and rotation about an axis perpendicular to the wall. For the full three-dimensional (3D) geometries, constructing a sufficiently refined mesh that contained a manageable number of elements proved difficult. We used local refinement with elements $1$ nm in size over the surface of the particle and in disks with radius $R$ on the bases of the cylinder around the cylinder's axis. The size of the elements was allowed to expand radially outward. We verified that integration over the surface yielded the surface area to within a fractional deviation of $10^{-5}$. We therefore expect that similar integration of the force and torque over the surface would show limited deviation between the true solution of the boundary value problem and our numerical result. To further demonstrate the quality of our approximate solution, we performed mesh refinement for a selected number of configurations and found no significant change.

It should be noted that there is no direct correspondence between the values we employed in our finite-element calculations and the parameters used in our LB simulations, since we are interested in fitting our simulations to the behavior expected for the Stokes' flow solution. We used reduced units throughout for our comparison between the raspberry particle simulations and the COMSOL calculations, therefore any specific choice of the parameters is factored out.

\subsection{\label{sub:notation}Notations Used throughout this Manuscript}

In this section, we summarize the notations used throughout this manuscript. This will aid in the understanding of our results, as many of the notations are necessarily similar.

\begin{itemize}
\setlength{\itemsep}{0pt}

\item $H$, the plate separation for the confined channel system, see Fig.~\ref{fig:plate}(a).

\item $R_{h}^{T}$, the effective hydrodynamic radius obtained by extrapolating translational mobility measurements, see Figs.~\ref{fig:plate}(a,b), for the limit of plate separation $H\uparrow\infty$. The subscript $h$ is used to differentiate $R_{h}$ from the raspberry radius $R$.

\item $R_{h}^{R}$, the effective hydrodynamic radius obtained by extrapolating rotational mobility measurements, see Fig.~\ref{fig:plate}(c), for the limit of plate separation $H\uparrow\infty$.

\item $z$, the vertical position of the sphere with respect to the center of the channel $(z=0)$, see Fig.~\ref{fig:plate}(a). 

\item $z_{\max}$, the maximum distance that a sphere can move from the center (z = 0) before making hydrodynamic contact with a wall: $z_{\max} = H/2 - R_{h}^{T}$ and $z_{\max} = H/2 - R_{h}^{R}$, respectively, see Fig.~\ref{fig:plate}(a).

\item $\mu^{T}_{0}$, the bulk translational mobility.

\item $\mu^{R}_{0}$, the bulk rotational mobility.

\item $\mu^{T}_{z,\perp}$, the translational mobility for movement along an axis perpendicular to the plates as a function of $z$ for fixed plate separation, see Fig.~\ref{fig:plate}(a).

\item $\mu^{T}_{z,\parallel}$, the translational mobility for movement along an axis parallel to the plates as a function of $z$ for fixed plate separation, see Fig.~\ref{fig:plate}(b).

\item $\mu^{R}_{z,\perp}$, the rotational mobility for rotation about an axis perpendicular to the plates as a function of $z$ for fixed plate separation, see Fig.~\ref{fig:plate}(c) left.

\item $\mu^{R}_{z,\parallel}$, the rotational mobility for rotation about an axis parallel to the plates as a function of $z$ for fixed plate separation, see Fig.~\ref{fig:plate}(c) right.

\item $\mu^{T}_{H,\perp}$, the translational mobility for movement along an axis perpendicular to the plates as a function of the plate separation $H$ for a sphere in the center of the channel $z=0$, see Fig.~\ref{fig:plate}(a).

\item $\mu^{T}_{H,\parallel}$, the translational mobility for movement along an axis parallel to the plates as a function of the plate separation $H$ for a sphere in the center of the channel $z=0$, see Fig.~\ref{fig:plate}(b).

\item $\mu^{R}_{H,\perp}$, the rotational mobility for rotation about an axis perpendicular to the plates as a function of the plate separation $H$ for a sphere in the center of the channel $z=0$, see Fig.~\ref{fig:plate}(c) left.

\item $\mu^{R}_{H,\parallel}$, the rotational mobility for rotation about an axis parallel to the plates as a function of the plate separation $H$ for a sphere in the center of the channel $z=0$, see Fig.~\ref{fig:plate}(c) right.

\item $f$, the fractional deviation between two results.

\end{itemize}

One final point of possible confusion is the use of the terms `parallel' and `perpendicular' to the plates. By parallel, we mean motion in a direction that is along the plates (orthogonal to the normal that defines the plates). By perpendicular motion, we mean in the direction parallel to this normal, \textit{i.e.}, away from one plate and towards the other. For rotation experiments these terms are used to describe the orientation of the axis of rotation with respect to the plates. Please refer to Fig.~\ref{fig:plate} and the above list if there is any confusion in reading the text.

\section{\label{sec:result}Results}

In this section, we discuss the results that we obtained by performing the simulations and numerical calculations outlined in Section~\ref{sec:methods}. 

\subsection{\label{sub:transplateperp}Perpendicular Translational Motion between Two Plates}

Let us begin by considering motion perpendicular to the plates, see Fig.~\ref{fig:plate}(a). This is the simplest case, since there are no translation-rotation cross-coupling terms in the HMT for this direction.~\cite{goldman67} The results shown in this section are for a filled raspberry model with radius $R = 3.0\sigma$ and $N_{\mathrm{tot}} = 925$ coupling points. However, we have also considered several other radii and the corresponding hollow equivalents of these raspberries. We will comment on this at the end of the section.

There are analytic expressions for the positional dependence $z$ of the perpendicular translational mobility $\mu^{T}_{z,\perp}$, to which we can compare our simulation results. These expressions follow from the result derived by Brenner for perpendicular movement near a single wall~\cite{brenner61} and were extended in Refs.~\cite{lobry96,lin00} to the geometry of two parallel plates. N.B. Both Refs.~\cite{lobry96,lin00} contain a typo in formulating the expression in Ref.~\cite{brenner61}, the correct expressions are given by
\begin{widetext}
\begin{eqnarray}
\label{eq:Brenner_alpha} \alpha & \equiv & \cosh^{-1}(p/R); \\
\label{eq:Brenner_plane} \lambda_{1}(p) & = & \frac{4}{3} \sinh(\alpha) \sum_{n=1}^{\infty} \left( \frac{n(n + 1)}{(2n - 1)(2n + 3)} \right) \left( \frac{ 2\sinh((2n + 1)\alpha) + (2n + 1)\sinh(2\alpha)}{4\sinh((n + 1/2)\alpha)^{2} - (2n + 1)^{2} \sinh(\alpha)^{2} } -  1 \right); \\
\label{eq:Lobry_Rice_corr} \lambda_{2}(z) & = & 1 + \sum_{n=0}^{\infty} \left[ \lambda_{1}\left( \frac{H}{2}(2n + 1) + z \right) - 1 \right] + \sum_{n=0}^{\infty} \left[ \lambda_{1}\left( \frac{H}{2}(2n - 1) - z \right) - 1 \right] - 2\sum_{n=0}^{\infty} \left[ \lambda_{1}\left( nH \right) - 1 \right] ; \\
\label{eq:mu_Brenner_plane} \mu^{T}_{\mathrm{s},\perp}(p) & = & \lambda_{1}^{-1}(p); \\
\label{eq:mu_Lobry_Rice_corr} \mu^{T}_{z,\perp} & = & \lambda_{2}^{-1}(z).
\end{eqnarray}
\end{widetext}
Here, $\mu^{T}_{\mathrm{s},\perp}(p)$ is the translational mobility of a sphere of radius $R$ moving perpendicular to a single wall. The variable $p$ is the minimal distance between the center of the sphere and this wall; the parameter $\alpha(p)$ is introduced to ease the notation. The perpendicular translational mobility between two parallel plates is obtained using $\mu^{T}_{\mathrm{s},\perp}(p)$ \textit{via} the image formalism. \textbf{N.B. There was a missing factor of $(4/3)\sinh(\alpha)$ in Eq.~\eqref{eq:Brenner_plane} in our J. Chem. Phys. paper, which has been corrected here.}

\begin{figure}[!htb]
\begin{center}
\includegraphics[scale=1.0]{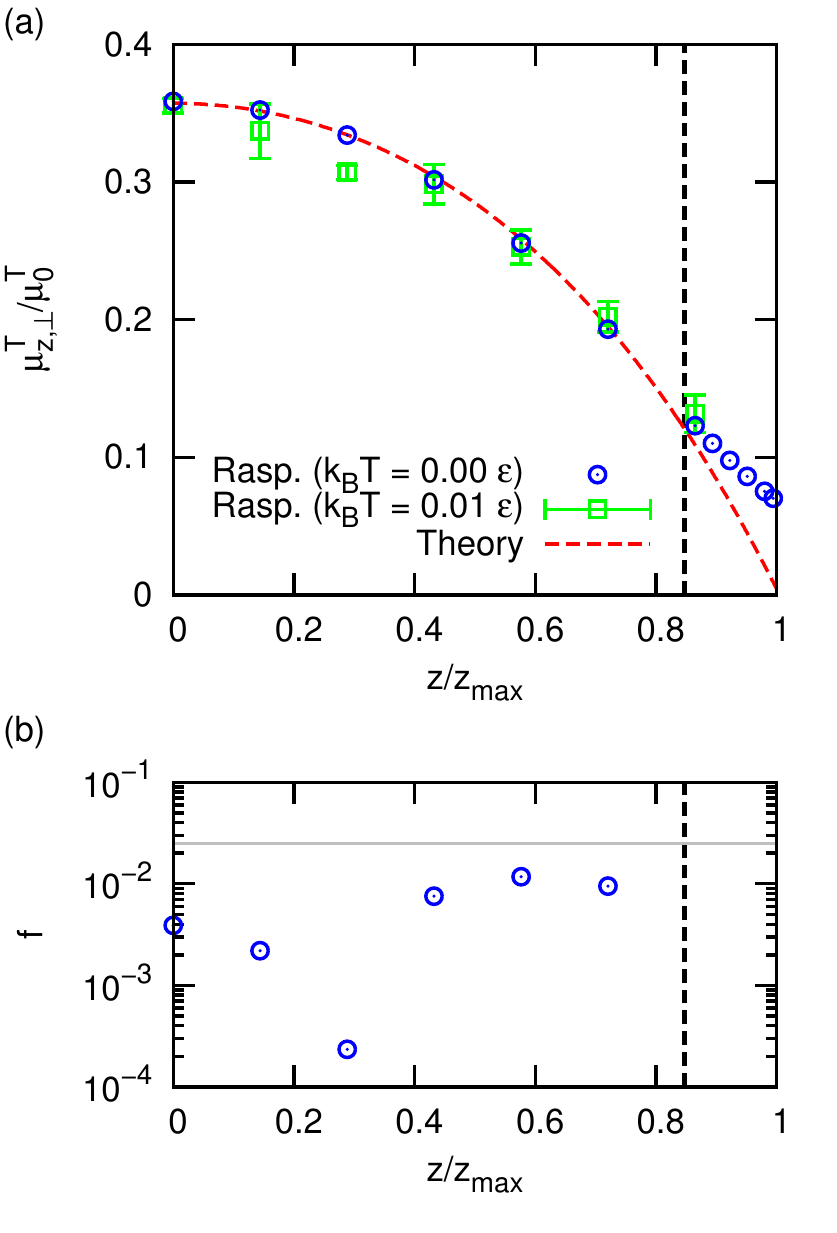}
\end{center}
\caption{\label{fig:perp_dist}(color online) The translational mobility perpendicular to the channel walls $\mu^{T}_{z,\perp}$ for a sphere of radius $R=3.0\sigma$ moving through a channel of height $H = 14.0\sigma$. The mobility is normalized by the bulk translational mobility $\mu^{T}_{0}$ and is given as a function of the position in the channel $z$, which itself is normalized by $z_{\max}$. The parameter $z_{\max}$ is the maximum distance that the sphere can move from the center ($z = 0$) before making contact with a wall, see the main text for further details. (a) The blue circles show the results of quiescent LB simulations, the green squares with error bars the results of thermalized LB simulations (the error bars indicate the standard error), and the dashed red curve the analytic result (Eq.~\cref{eq:mu_Lobry_Rice_corr} and Refs.~\cite{brenner61,lobry96,lin00}). The dashed vertical line (thick black) indicates the position for which the sphere and the wall are separated by one LB lattice spacing. (b) The fractional deviation $f$ of the quiescent LB result from the theoretical result. The horizontal gray line shows a fractional deviation of $2.5\%$.}
\end{figure}

Figure~\ref{fig:perp_dist}(a) shows the result of our quiescent velocity and thermalized MSD experiments, Figs.~\ref{fig:plate}(a)~and~\ref{fig:plate}(d), respectively, compared to the analytic value of $\mu^{T}_{z,\perp}$, see Eq.~\cref{eq:mu_Lobry_Rice_corr}. Here, we used a channel height of $H = 14.0\sigma$, in which we placed a filled raspberry with radius $R=3.0\sigma$. As can be seen in Fig.~\ref{fig:perp_dist}(b) the agreement between our quiescent simulations and the corrected series expression is quite excellent. There is a fractional deviation $f$ of less than $1\%$ throughout the channel, up to the point where the raspberry and the channel walls are separated by less than $1.0\sigma$. Here it should be noted that the mapping of the position to $z/z_{\max}$ is based on $R^{T}_{h}$. We will come back to the effective hydrodynamic radius $R^{T}_{h}$, by which we calculated $z_{\max}$, and the bulk mobility by which we normalize $\mu^{T}_{z,\perp}$ shortly.

For small separations there is a systematic deviation between the theory and our LB results, see Fig.~\ref{fig:perp_dist}(a). This deviation can be attributed to the break-down of the description in terms of an effective hydrodynamic radius in proximity of a wall (within one lattice spacing).  Moreover, the lubrication forces are not accurately accounted for in our combined MD-LB model for close wall-particle proximity. This could explain why $\mu^{T}_{z,\perp} \ne 0$ for $z = z_{\max}$. Finally, note that the results of our thermalized MSD experiment match the trend of the theoretical curve as well, within the error bar. 

\begin{figure}[!htb]
\begin{center}
\includegraphics[scale=1.0]{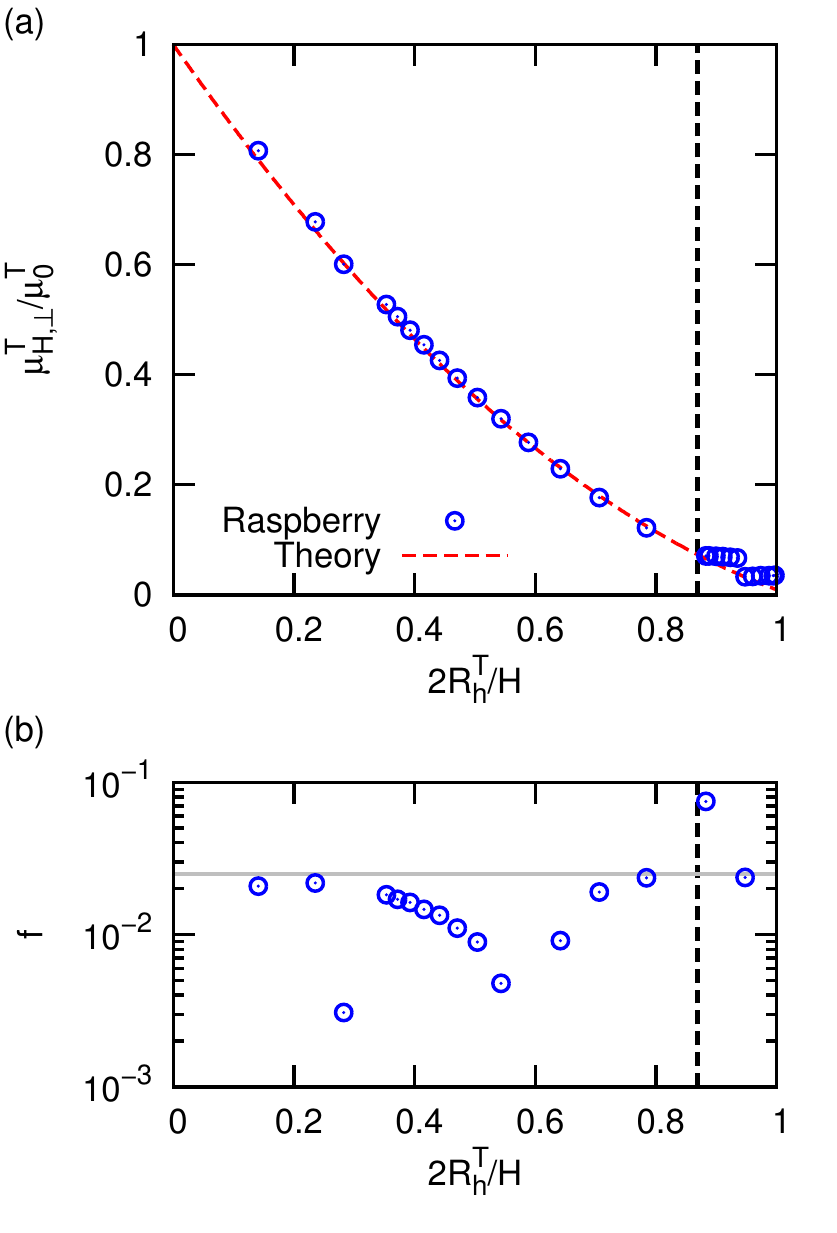}
\end{center}
\caption{\label{fig:perp_height}(color online) The translational mobility perpendicular to the channel walls $\mu^{T}_{H,\perp}$ for a sphere of radius $R=3.0\sigma$ moving through a channel of height $H$ in the center of the channel $z=0$. The mobility is normalized by the bulk translational mobility $\mu^{T}_{0}$ and is given as a function of the reciprocal of the channel height $H$, multiplied by the hydrodynamic diameter (twice the hydrodynamic radius $R^{T}_{h}$). (a) The blue circles show the results of quiescent LB simulations and the dashed red curve the semi-analytic result.~\cite{brenner61,lobry96,lin00} The dashed vertical line (thick black) indicates the position for which the sphere (which is in the center of the channel) and the two walls are separated by one LB lattice spacing. (b) The fractional deviation $f$ of the quiescent LB data from the theoretical result. The horizontal gray line shows a fractional deviation of $2.5\%$.}
\end{figure}

In order to determine the value of the effective hydrodynamic radius, we varied the distance between the plates and measured $\mu^{T}_{z,\perp}$ at $z=0$ (the middle of the channel), where the mobility attains its highest value. We denote this plate-separation dependent mobility by $\mu^{T}_{H,\perp}$. The result of our simulations, compared to the theoretical expression in Eq.~\cref{eq:mu_Lobry_Rice_corr} (setting $z = 0$ and varying $H$) is shown in Fig.~\ref{fig:perp_height}. Agreement between the two data sets is quite excellent for $H > 2R^{T}_{h}+\sigma$, as can be appreciated from Fig.~\ref{fig:perp_height}(b), which shows systematic deviation between the theory and numerical results, but it is small. The structure in $f$ originates from the fitting procedure used to extract the bulk mobility and its subsequent use as a normalization factor. Furthermore, the model breaks down for smaller separations, as can be clearly seen in Fig.~\ref{fig:perp_height}(a), since sub-lattice changes in the plate separation cannot be obtained using the bounce-back boundaries.

Finally, repeating the same experiments with the hollow model reveals that the effective radii that we obtained from both models were close to the ones that we obtained from our experiments in the simple-cubic geometry, which we discussed in Part I.~\cite{fischer15} This is shown in Table~\ref{tab:sum} and will be discussed in detail in Section~\ref{sec:disc}. We also performed plate separation simulations for our raspberries with $R = 2.0\sigma$, $2.5\sigma$, $3.0\sigma$, $4.0\sigma$, and $5.0\sigma$. For both the hollow and the filled raspberry model and all sizes, the dependence of $\mu^{T}_{z,\perp}$ and $\mu^{T}_{H,\perp}$ on $z$ and $H$, respectively, is the same within the error bar (not shown here). 

\subsection{\label{sub:transplatepara}Parallel Translational Motion between Two Plates}

We continued our investigation into the quality of the raspberry model under confinement by examining the mobility in the direction parallel to the plates, see Fig.~\ref{fig:plate}(b). That is, motion perpendicular to the normal vector that defines the plate orientation. For such motion there are translation-rotation cross-coupling terms in the HMT.~\cite{goldman67} Only when the sphere moves exactly in the plane centered between the two plates, there is no coupling due to symmetry.

\begin{figure}[!htb]
\begin{center}
\includegraphics[scale=1.0]{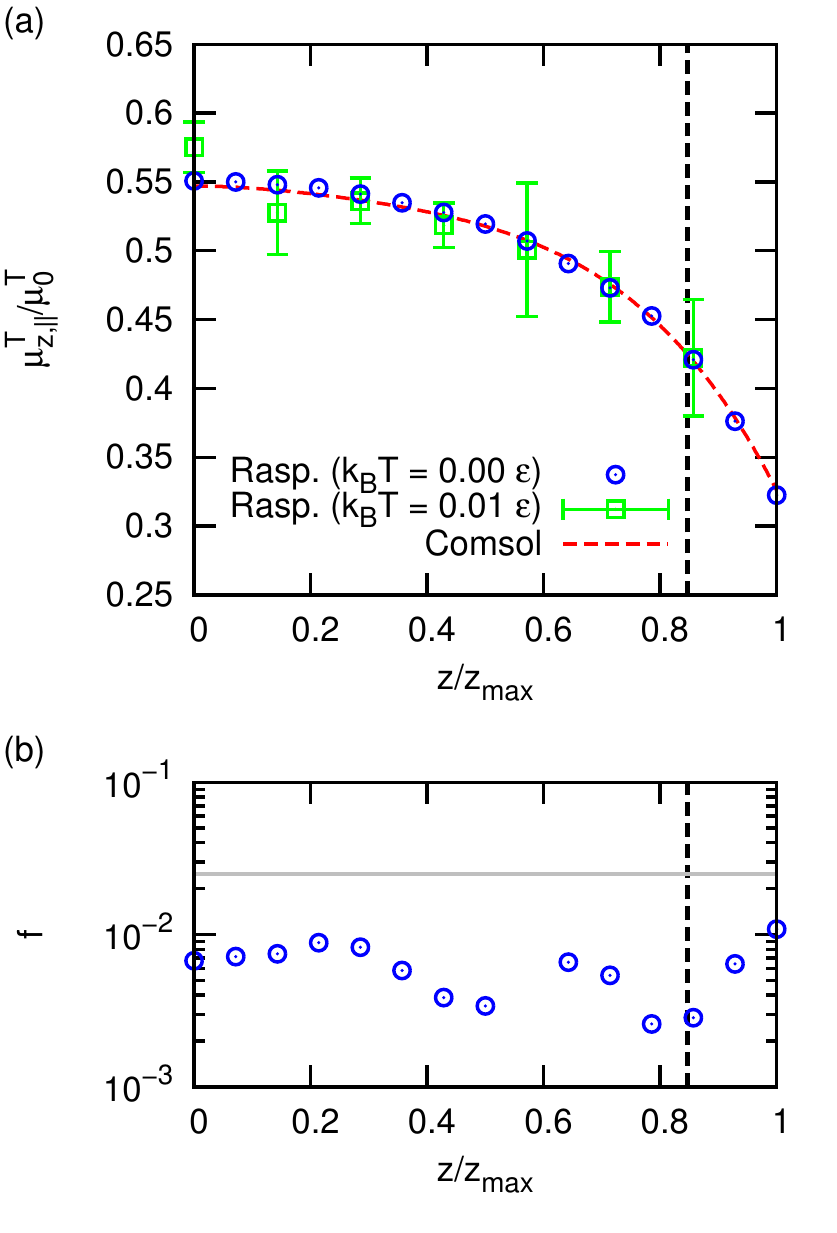}
\end{center}
\caption{\label{fig:para_dist}(color online) The translational mobility parallel to the channel walls $\mu^{T}_{z,\parallel}$ for a sphere ($R=3.0\sigma$) moving through a channel of height $H = 14.0\sigma$, plotted as a function of the position $z$. The notation is otherwise the same as in Fig.~\ref{fig:perp_dist}. (a) Quiescent LB simulations (blue circles), thermalized LB simulations (green squares with error bars), and COMSOL results (dashed red curve). (b) The fractional deviation $f$ between the quiescent LB and COMSOL result (gray line $f=2.5\%$).}
\end{figure}

Figure~\ref{fig:para_dist}(a) shows the result of our quiescent velocity (Fig.~\ref{fig:plate}(b)) and thermalized MSD experiments (Fig.~\ref{fig:plate}(d)) compared to the $\mu^{T}_{z,\parallel}$ obtained using COMSOL. The COMSOL results were fitted with a fourth order polynomial. Here, we used a channel height of $H = 14.0\sigma$ in which we placed a filled raspberry with radius $R=3.0\sigma$. As can be seen in Fig.~\ref{fig:para_dist}(c) the agreement between our quiescent simulations and the finite-element calculations is quite excellent, despite the fact that we ignored cross-coupling in the latter. The fractional deviation $f$ is less than $1\%$ throughout the channel, even for points where the raspberry and the channel walls are separated by less than $1.0\sigma$. The results of our thermalized MSD experiment match the trend of the finite-element curve as well, within the error bar, see Fig.~\ref{fig:para_dist}(a). We should note that in our simulations we found no evidence of translation-rotation cross-coupling. This is presumably due to the small particle-wall separations, for which these effects typically express themselves.~\cite{goldman67}

For motion in the center plane there is an analytic expression for height dependence of the parallel translational mobility $\mu^{T}_{H,\parallel}$ derived by Faxen~\cite{faxen22,happel83}. Faxen's law for center-plane parallel motion between two plates is given by
\begin{eqnarray}
\nonumber \frac{\mu^{T}_{H,\parallel}}{\mu^{T}_{0}} & = & 1.000 - 1.004x + 0.418x^{3} + 0.210x^{4} - 0.169x^{5} ; \\
\label{eq:Faxen} & & \\
\label{eq:x} x & \equiv & \frac{2R^{T}_{h}}{H} ,
\end{eqnarray}
It is claimed that Eq.~\cref{eq:Faxen} is accurate up to fifth order in the plate separation.~\cite{happel83} However, the fifth order coefficient was only provided in a private communication~\cite{happel83} and no account of its derivation exist. 

\begin{figure}[!htb]
\begin{center}
\includegraphics[scale=1.0]{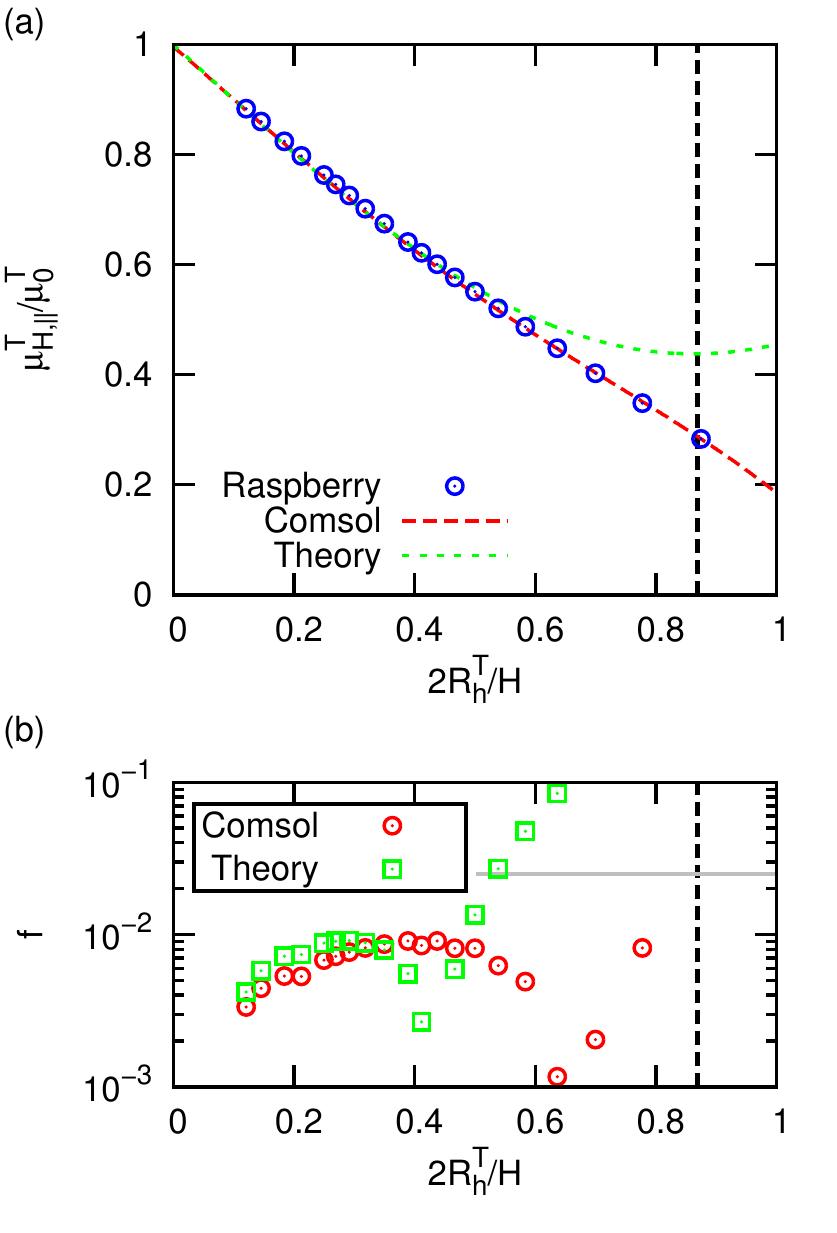}
\end{center}
\caption{\label{fig:para_height}(color online) The translational mobility parallel to the channel walls in the center of the channel $\mu^{T}_{H,\parallel}$ for a sphere of radius $R=3.0\sigma$ moving through a channel of height $H$. The notation is otherwise the same as in Fig.~\ref{fig:perp_height}. (a) The blue circles show the data from quiescent LB simulations, the dashed green curve the theoretical approximation by Faxen (Eq.~\cref{eq:Faxen}; Refs.~\cite{faxen22,happel83}), and the dashed red curve the result of finite-element calculations performed using COMSOL (Eq.~\cref{eq:Faxmod}). (b) The fractional deviation $f$ of the quiescent LB result from the theoretical result by Faxen (green squares) and the COMSOL result (red circles). The horizontal gray line shows a fractional deviation of $2.5\%$.}
\end{figure}

To determine the bulk mobility and the associated effective hydrodynamic radius, as well as examine the quality of Faxen's law, we examined $\mu^{T}_{H,\parallel}$ for parallel motion in the center plane. The result of our simulations is shown in Fig.~\ref{fig:para_height}, which provides a comparison between our data, the theoretical result by Faxen (Eq.~\cref{eq:Faxen}),~\cite{faxen22,happel83} and the COMSOL finite-element calculations. The result obtained using the raspberry model agrees perfectly with the finite-element data over the entire range of plate separations considered here. As expected there is a significant deviation between Faxen's result for small plate separations, since the theory ignores higher-order contributions in $H$ to the mobility. We only find acceptable agreement between our data and Faxen's expression for $H \gtrsim 4R^{T}_{h}$. 

From our finite-element calculations we can establish an `improved' version of Faxen's law. The COMSOL data points in Fig.~\ref{fig:para_height}(a) were fitted using a fourth-order polynomial
\begin{eqnarray}
\label{eq:Faxmod} \frac{\mu^{T}_{H,\parallel}}{\mu^{T}_{0}} & = & A + Bx + Cx^{2} + Dx^{3} + Ex^{4}; \\
\label{coefA} A & =      & +0.997\pm0.001 ; \\
\label{coefB} B & =      & -0.949\pm0.009 ; \\
\label{coefC} C & =      & -0.215\pm0.034 ; \\
\label{coefD} D & =      & +0.895\pm0.051 ; \\
\label{coefE} E & =      & -0.542\pm0.025 .
\end{eqnarray}
The numbers with error bars give our numerically obtained coefficients. \textbf{N.B. The coefficients here are correct, for the ones listed in our J. Chem. Phys. publication the coefficients $D$ and $E$ are reversed.} Note that we used a different form of polynomial expansion for our improved Faxen's law (leaving in the $x^{2}$ term) to achieve a compacter expression and better $\chi^{2}$ values, since we fit lower-order terms. The first coefficient $A$ can be taken to be $1$ without significantly changing the result of the fitting procedure. This would give the correct limit for $x \downarrow 0$, which corresponds to infinite plate separation.

We again found for both the hollow and the filled raspberry, as well as all raspberry radii which we investigated, that the dependence of $\mu^{T}_{z,\parallel}$ and $\mu^{T}_{H,\parallel}$ on $z$ and $H$, respectively, is the same within the error bar (not shown here). The value of the effective radii for the hollow and filled models differed, though, see Table~\ref{tab:sum}. This is in agreement with our findings for the perpendicular motion between plates and in the simple-cubic geometry.~\cite{fischer15} We will discuss this further in Section~\ref{sec:disc}.

\subsection{\label{sub:rotplate}Rotational Motion between Two Plates}

Finally, we considered the accuracy of the raspberry model under confinement for the case of rotation about an axis in the direction parallel and perpendicular to the plates, see Fig.~\ref{fig:plate}(c). For rotation about an axis parallel to the wall (perpendicular to the normal, see Fig.~\ref{fig:plate}(c) (left)) there are rotation-translation cross-coupling terms in the HMT.~\cite{goldman67} Only when $z=0$, there is no such coupling due to symmetry considerations. To the best of our knowledge no analytic expressions for rotational mobilities between two plates exist. Therefore, we compare to finite-element calculations, the data points of which we fitted using a fourth-order polynomial. We only show results for the filled raspberry model for the position $z$ and height $H$ dependence of the mobilities.

\begin{figure}[!htb]
\begin{center}
\includegraphics[scale=1.0]{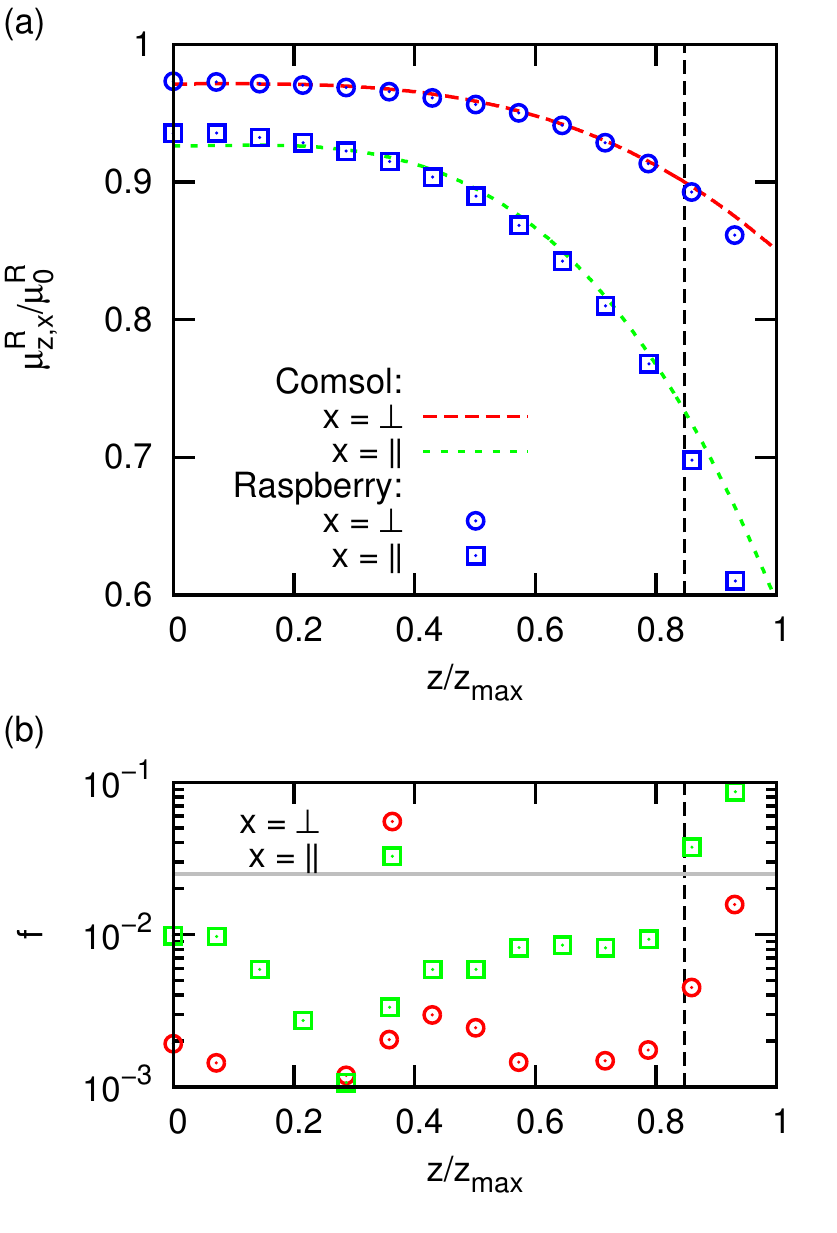}
\end{center}
\caption{\label{fig:rot_dist}(color online) The rotational mobility for rotation about an axis perpendicular and parallel to the walls ($\mu^{R}_{z,\perp}$ and $\mu^{R}_{z,\parallel}$, respectively) for a sphere of radius $R=3.0\sigma$ moving through a channel of height $H = 14.0\sigma$. The mobilities are normalized by the bulk rotational mobility $\mu^{R}_{0}$ and plotted as a function of the position $z$ from the center $(z=0)$, expressed in terms of the maximal deviation $z_{\max}$. The notation is otherwise the same as in Figs.~\ref{fig:perp_dist} and~\ref{fig:para_dist}. (a) The blue circles show $\mu^{R}_{z,\perp}$ and the blue squares show $\mu^{R}_{z,\parallel}$ for quiescent LB simulations, the dashed red and green curves show the corresponding results obtained using COMSOL. (b) The fractional deviation $f$ between the quiescent LB and the COMSOL result, rotation about an axis perpendicular to the walls is indicated using red circles and rotation about an axis parallel to the wall using green squares (gray line $f=2.5\%$).}
\end{figure}

Figure~\ref{fig:rot_dist}(a) shows the result of our quiescent torque experiments compared to the data obtained using COMSOL for both forms of rotation. Here, we used a channel height of $H = 14.0\sigma$ in which we placed a filled raspberry with radius $R=3.0\sigma$. As can be seen in Fig.~\ref{fig:rot_dist}(b) the agreement between our quiescent simulations and the finite-element calculations is quite excellent. The fractional deviation $f$ is less than $1\%$ throughout the channel for rotation about an axis parallel to the wall and even better for rotation about an axis perpendicular to the wall. We can see significant deviation for the former when the raspberry and the channel walls are separated by less than $1.0\sigma$, but this can be explained by the small particle-wall separation compared to the lattice spacing. Again, we did not observe translation-rotation cross coupling in our data. 

Note that there is some structure to the deviation between the parallel-rotation raspberry data and the finite-element calculations. This is related to the fitting procedure used to extract the bulk rotational mobility $\mu^{R}_{0}$ by varying $H$. A small deviation in the effective hydrodynamic radius $R^{R}_{h}$ can have significant impact on the shape of the curve, as we will explain shortly. For the rotation about an axis perpendicular to the wall there is a similar increase when the wall-raspberry separation is less than $1.0\sigma$, but the agreement is still quite reasonable. 

\begin{figure}[!htb]
\begin{center}
\includegraphics[scale=1.0]{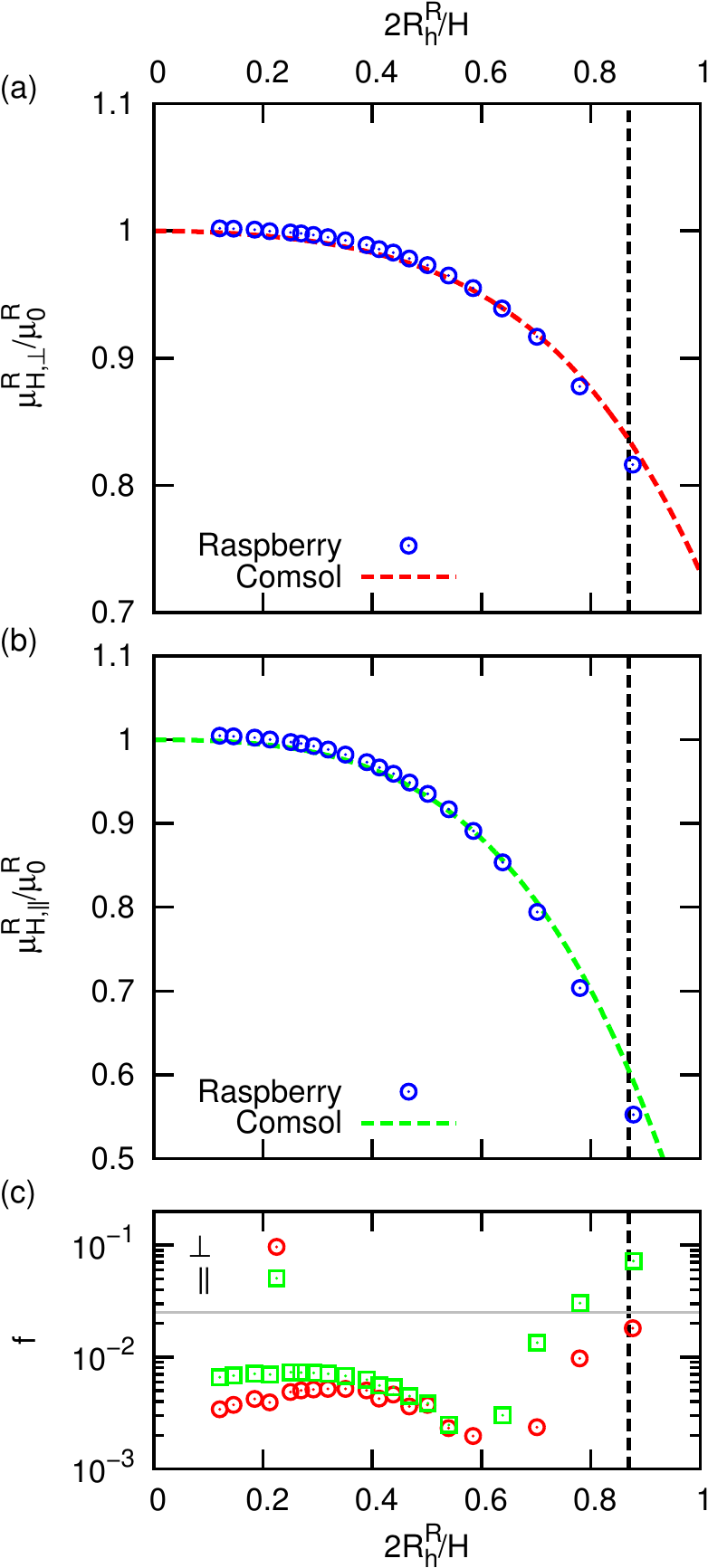}
\end{center}
\caption{\label{fig:rot_height}(color online) The perpendicular (a) and parallel (b) rotational mobility ($\mu^{R}_{z,\perp}$ and $\mu^{R}_{z,\parallel}$, respectively) for a sphere of radius $R=3.0\sigma$ moving through a channel of height $H$. The mobilities are  normalized by the bulk coefficient $\mu^{R}_{0}$ and plotted as a function of $1/H$, expressed in terms of twice the hydrodynamic radius $R^{R}_{h}$. The notation is otherwise the same as in Figs.~\ref{fig:perp_height} and~\ref{fig:para_height}. Quiescent LB simulations are given by blue circles and COMSOL results using dashed red curves. (c) The fractional deviation $f$ of the quiescent LB results from the COMSOL data (gray line $f=2.5\%$).}
\end{figure}

To determine the bulk mobility and the associated effective hydrodynamic radius, we examined $\mu^{R}_{H,\perp}$ and $\mu^{R}_{H,\parallel}$ (rotation in the mid-plane with varying $H$), see Fig~\ref{fig:rot_height}. Again we find quite acceptable agreement between the data obtained using the raspberry model and the finite-element calculations. However, we can clearly see that there is structure to the fractional deviation $f$ between the two results in Fig.~\ref{fig:rot_height}(c). This can be explained as follows.

In determining the effective hydrodynamic radius, we used a quartic polynomial to fit our data, which was given as a function of $2R^{R}_{h}/H$ and normalized by the value of $\mu^{R}_{0}$ corresponding to that $R^{R}_{h}$. That is, the data to which we fitted is a function of $R^{R}_{h}$ and we applied a root-finding algorithm to ensure that the fitted value for $H \uparrow \infty$ of $\mu^{R}_{H,\perp}/\mu^{R}_{0}$ and $\mu^{R}_{H,\parallel}/\mu^{R}_{0}$ equals $1$. We thus solved for the value $R^{R}_{h}$. 

The above procedure, however, strongly favors the criterion imposed on the extrapolated data point, while it sacrifices overall matching to the finite-element calculations -- which we assume to be more accurate. Hence, there can be a systematic deviation between the two data sets, as is indeed found in Fig.~\ref{fig:rot_height}(c). However, our fitting procedure is reasonable, since it does not implicitly assumes the finite-element data to be superior to that obtained using the raspberry model and it only uses internal fitting parameters. The agreement between the two results, despite the systematic deviations, can therefore be considered excellent.

From our numerical results we can formulate modified empirical forms for the rotational mobility's dependence on the channel height $H$, for mid-channel rotation about an axis parallel and perpendicular to the walls. These expressions are as follows
\begin{eqnarray}
\label{eq:Faxrot} \frac{\mu^{R}_{H,i}}{\mu^{R}_{0}} & = & 1 + A_{i}x^{2} + B_{i}x^{3} + C_{i}x^{4}; \\
\label{eq:xrot} x & \equiv & \frac{2R^{R}_{h}}{H} ;\\
\label{coefArot} A_{\perp} & = & -0.10\pm0.01 ; \quad A_{\parallel} = -0.09\pm0.01 ;\\
\label{coefBrot} B_{\perp} & = & +0.09\pm0.03 ; \quad B_{\parallel} = -0.17\pm0.04 ;\\
\label{coefCrot} C_{\perp} & = & -0.26\pm0.02 ; \quad C_{\parallel} = -0.38\pm0.03 ,
\end{eqnarray}
where $i$ is a subscript that can be either $\perp$ for rotation about an axis perpendicular to the walls and $\parallel$ for rotation about an axis parallel to the walls. We divided out the linear coefficient $\mu_{0}^{R}$ to obtain the functional form of our result.

Again, we found that for both the hollow and the filled raspberry model, as well as all three raspberry radii which we studied, the dependence of $\mu^{R}_{z,\perp}$, $\mu^{R}_{z,\parallel}$, $\mu^{R}_{H,\perp}$, and $\mu^{R}_{H,\parallel}$ on $z$ and $H$, respectively, is the same within the error bar (not shown here). The value of the effective radii for the hollow and filled models differed in each case, see Table~\ref{tab:sum}, in agreement with our findings for the perpendicular and parallel motion between plates, as well as in the simple-cubic geometry.~\cite{fischer15} We will discuss this further in Section~\ref{sec:disc}.

\section{\label{sec:disc}Discussion}

In Section~\ref{sec:result} we have shown that there is excellent agreement between established analytical and numerical results for the Stokesian hydrodynamic behavior of spherical particles compared to their raspberry equivalent in the confining geometry of two parallel plates. This extends our findings of Part I~\cite{fischer15} to geometries with walls. The discrepancy between the translational and rotational hydrodynamic radius of the original hollow raspberry -- first observed by Ollila~\textit{et al.}~\cite{ollila13} -- is present for the parallel plate system as well. In this section we discuss this discrepancy in more detail.

\begin{table}
\begin{ruledtabular}
\begin{tabular}{c|c|c|c|c}
                  & \multicolumn{2}{c|}{filled}  & \multicolumn{2}{c}{hollow}  \\
\hline
\hline
\\[-0.8em]
$R = 2.0\sigma$   & $R^{T}_{h}$  & $R^{R}_{h}$  & $R^{T}_{h}$  & $R^{R}_{h}$  \\
\hline
\\[-0.8em]
simple-cubic      & $2.59\sigma$ & $2.57\sigma$ & $2.50\sigma$ & $2.42\sigma$ \\
walls $\perp$     & $2.57\sigma$ & $2.56\sigma$ & $2.48\sigma$ & $2.43\sigma$ \\
walls $\parallel$ & $2.57\sigma$ & $2.56\sigma$ & $2.48\sigma$ & $2.43\sigma$ \\
\hline
\hline
\\[-0.8em]
$R = 2.5\sigma$   & $R^{T}_{h}$  & $R^{R}_{h}$  & $R^{T}_{h}$  & $R^{R}_{h}$  \\
\hline
\\[-0.8em]
simple-cubic      & $3.01\sigma$ & $3.02\sigma$ & $2.98\sigma$ & $2.90\sigma$ \\
walls $\perp$     & $3.00\sigma$ & $3.01\sigma$ & $2.97\sigma$ & $2.90\sigma$ \\
walls $\parallel$ & $3.00\sigma$ & $3.01\sigma$ & $2.97\sigma$ & $2.90\sigma$ \\
\hline
\hline
\\[-0.8em]
$R = 3.0\sigma$   & $R^{T}_{h}$  & $R^{R}_{h}$  & $R^{T}_{h}$  & $R^{R}_{h}$  \\
\hline
\\[-0.8em]
simple-cubic      & $3.53\sigma$ & $3.54\sigma$ & $3.47\sigma$ & $3.38\sigma$ \\
walls $\perp$     & $3.53\sigma$ & $3.50\sigma$ & $3.48\sigma$ & $3.44\sigma$ \\
walls $\parallel$ & $3.53\sigma$ & $3.51\sigma$ & $3.47\sigma$ & $3.37\sigma$ \\
\hline
\hline
\\[-0.8em]
$R = 4.0\sigma$   & $R^{T}_{h}$  & $R^{R}_{h}$  & $R^{T}_{h}$  & $R^{R}_{h}$  \\
\hline
\\[-0.8em]
simple-cubic      & $4.57\sigma$ & $4.56\sigma$ & $4.46\sigma$ & $4.37\sigma$ \\
walls $\perp$     & $4.55\sigma$ & $4.54\sigma$ & $4.45\sigma$ & $4.36\sigma$ \\
walls $\parallel$ & $4.54\sigma$ & $4.54\sigma$ & $4.44\sigma$ & $4.36\sigma$ \\
\hline
\hline
\\[-0.8em]
$R = 5.0\sigma$   & $R^{T}_{h}$  & $R^{R}_{h}$  & $R^{T}_{h}$  & $R^{R}_{h}$  \\
\hline
\\[-0.8em]
simple-cubic      & $5.55\sigma$ & $5.53\sigma$ & $5.44\sigma$ & $5.35\sigma$ \\
walls $\perp$     & $5.55\sigma$ & $5.52\sigma$ & $5.44\sigma$ & $5.35\sigma$ \\
walls $\parallel$ & $5.54\sigma$ & $5.52\sigma$ & $5.43\sigma$ & $5.35\sigma$ \\
\end{tabular}
\end{ruledtabular}
\caption{\label{tab:sum}Summary of the various effective radii determined in the experiments described in Section~\ref{sec:result} and Ref.~\cite{fischer15}. There is a $\pm0.01\sigma$ error bar for the translational experiments and $\pm 0.02\sigma$ for the rotational experiments. The bare friction coefficient used to generate this data is $\zeta_{0} = 25 m_{0} \tau^{-1}$.}
\end{table}

In Table~\ref{tab:sum} we summarize the results for the hydrodynamic radii obtained in our experiments in this manuscript as well as in Part I.~\cite{fischer15} Note that within the error bar the data for the filled raspberries are fully consistent, whereas the data from the hollow raspberries show a clear discrepancy between $R^{T}_{h}$ and $R^{R}_{h}$. The $R^{R}_{h}$ obtained from the experiment ($R = 3.0\sigma$) with rotation about an axis perpendicular to the wall (Fig.~\ref{fig:plate}(c), right) seems to be an outlier for both the filled and hollow raspberry, but it is still within two standard deviations of the mean for the filled raspberry. As explained in Section~\ref{sub:rotplate}, our fitting procedure ensured that for the box length $L \uparrow \infty$ in the simple-cubic geometry~\cite{fischer15} and for the plate separation $H \uparrow \infty$ in the confined geometry, the fitted result converges to the expected bulk mobility for a given hydrodynamic radius, for which we subsequently solved. Therefore, the rotational data has a higher standard deviation in general, as taking the cubic root to establish the effective radius leads to stronger error propagation. 

From our results, we conclude that the level of disagreement between the effective radii of the hollow raspberry model seems to be unaffected by the confinement. This can be seen from the excellent agreement (in an absolute sense) between the hydrodynamic radii in bulk and good relative agreement between our results and literature references in the confining geometry. We have also shown that our `filling + fitting' procedure for the raspberry model (see Ref.~\cite{fischer15}) again significantly improves the agreement between the effective hydrodynamic radius obtained by translational and rotational experiments. There is correspondence between the effective radii that are found for the `filling + fitting' raspberry using the various experiments, within the numerical error. A similar conclusion cannot be reached for the hollow raspberry, indicating that this model may be more sensitive to the details of its surroundings.

Finally, let us focus on the raspberry constructed using the Ahlrichs and D{\"u}nweg viscous coupling (VC) scheme~\cite{ahlrichs99} in the context of the other methods available to model the hydrodynamic interactions of colloids with each other and boundaries. There are several well-known alternatives to the method outlined in Part I~\cite{fischer15} and this manuscript, by which MD objects can be coupled to LB, namely: Ladd bounce-back boundaries (Ladd BB),~\cite{frisch87,Ladd94} the immersed boundary method (IBM),~\cite{Peskin02} and the external boundary force (EBF) method.~\cite{Wu10} There are other hydrodynamics solvers that can achieve similar colloid-in-fluid descriptions. Most commonly used in physics are dissipative particle dynamics (DPD),~\cite{Hoogerbrugge92,Espanol95} multi-particle collision dynamics (MPCD) or stochastic rotation dynamics (SRD),~\cite{Malevanets99,Ihle03} and Stokesian Dynamics (SD).~\cite{brady88} In the following we briefly discuss our method in the context of these alternatives.

It was recently shown by Schiller~\cite{schiller14} that VC is a general description for fluid-particle coupling, which encompasses both IBM and EBF. IBM and EBF correspond to special choices of the friction and mass ratio in VC. As such, for the LB-based descriptions it suffices to consider Ladd BB. One point of interest is that the EBF method can be modified to lead to an instantaneous no-slip boundary condition, but only at the expense of loosing freedom in the choice of the bare friction coefficient.~\cite{schiller14}

Ladd BB employs the grid on which the LB fluid is simulated to describe the particles.~\cite{frisch87,Ladd94} This gives Ladd BB an advantage over the VC raspberry model, as it does not necessitate an interpolation step or sub-lattice refinement to achieve the fluid-particle coupling. Similar to the VC raspberry particle, Ladd BB also gives rise to an effective hydrodynamic radius, for which can be fitted. However, Ladd BB achieves a no-slip boundary condition, without necessitating a particular choice of the bare friction coefficient. This is favorable, as there is no question of porosity.~\cite{Ollila12,ollila13,fischer15} It is therefore likely that Ladd BB performs better in force and velocity experiments of Part I,~\cite{fischer15} as no counter-force can be applied to the nodes which are encompassed by the Ladd BB particle, leading to an effective (reduced) force being applied to the particle. Finally, there are lubrication corrections available for the Ladd BB method.~\cite{Ladd01,Ding03} These should improve the near-field interactions between particles and particles and walls, compared to the VC raspberry method. 

Considering these advantages Ladd BB would seem the method of choice. However, the grid-based description of the particle in Ladd BB is known to suffer when there are too few grid points inside the particle. For example, to obtain a coupling shape that is sufficiently spherical, a radius of $R \ge 2.5\sigma$ is required. We have shown that the VC raspberry can be used for smaller $R$ and the VC has been shown to work well even for single coupling points.~\cite{ahlrichs99} This is useful in systems where one is restricted in the number of LB nodes that one can use -- in particular graphics processor unit (GPU)-based LB codes are limited by the amount of RAM available on the GPU. VC allowes one to study large systems with many thousands of particles, without the need to fully resolve each on the lattice, see, \textit{e.g.}, Ref.~\cite{roehm14}. 

The non-LB methods to achieve fluid-colloid coupling vary considerably in the way that this coupling is achieved. DPD, MPCD, and SRD (the latter two essentially being the same), all achieve coupling to the `fluid' by interacting with the particles that comprise it. For MPCD and SRD, the smallest length scale on which the larger raspberry object can be resolved is governed by the size of the fluid particles, rather than the boxes that are used in the collision step. This could well be advantageous in modeling small separations of walls and raspberries. However, it should be noted that a sufficient density of MPCD particles must be present to properly model hydrodynamic interactions, which would be problematic for situations in which the raspberries are in proximity to each other and obstacles. In DPD, however, the finest features of the raspberry that can be resolved have the typical length scale of the solvent particle interactions. As such, DPD is typically a coarser method than both the VC in LB and MPCD and is more likely to encounter problems for small raspberry-raspberry and raspberry-wall separations. 

Finally, there is the method of SD,~\cite{brady88} in which the hydrodynamic interactions between particles are described using the Rotne-Prager tensor. This method a suited for bulk systems, but it can be modified to periodic boundary conditions. Its main attraction is the fact that it solves Stokes' equations exactly for dilute systems and that it can be modified to include a lubrication correction. This together with the algorithm's favorable scaling~\cite{brady01} give it a slight edge over the VC method for bulk systems. However, simple (closed) analytic forms for the Rotne-Prager tensor are not available for more complicated geometries -- not even for two plates. Only for single plates can one use the Blake-tensor formalism. Therefore, any advantage of the SD algorithm in bulk (with or without lubrication corrections) is lost in more complicated geometries, since the quality of the solution sensitively depends on the approximations made for the hydrodynamic interaction tensors.

A full comparison of the various methods goes beyond the scope of the current research, but is certainly worthwhile for future study.

\section{\label{sec:conc}Conclusion and Outlook}

Summarizing, we have examined the properties of the raspberry model using a classic fluid dynamics experiments, wherein a sphere is subjected to the confining geometry of parallel plates. This so-called `raspberry' model refers to a hybrid lattice-Boltzmann (LB) and Langevin molecular dynamics (MD) scheme for simulating the dynamics of suspensions of colloids originally developed by Lobaskin and D{\"u}nweg.~\cite{lobaskin04} The particle is represented by a set of points on its surface that couple to the fluid through a frictional force acting both on the solvent and on the solute, which depends on the relative velocity. 

We began our investigation by analyzing the position dependent translational mobility of a sphere moving perpendicular to the two plates. We found excellent agreement with the predictions of Refs.~\cite{brenner61,lobry96,lin00}. This was extended upon by considering the parallel translational mobility of the sphere. Here, we found good agreement with the result by Faxen~\cite{faxen22,happel83} in the limit of large plate separation. We improved upon Faxen's result by proposing a modified, empirical expression that works well over the entire range of plate separations. Finally, we considered the rotational mobility for rotation about axes parallel and perpendicular to the plates. For these systems no analytic expressions were available to which we could compare our numerical results. However, on the basis of our numerical results, we were able to formulate expressions for these mobility coefficients for spheres fixed in the center of the plates as a function of the plate separation. 

Our results show that Stokesian hydrodynamic behavior is reproduced by our raspberry model with the `filling + fitting' formalism to a surprising degree of accuracy over a wide range of length scales. From our combined data we can draw the following additional conclusions concerning our `filling + fitting' procedure and the raspberry point-coupling method in general:
\begin{itemize}
\item To determine the bulk translational mobility of a particle, a force or velocity experiment performed in the center of two plates is more suitable than in a simple-cubic geometry. This type of experiment does not have the disadvantage of a back force/velocity density,~\cite{fischer15} which could interfere with the fitting procedure required to extrapolate results to bulk (infinite plate separation).
\item The raspberry point-coupling can be combined with confining geometries consisting of bounce-back boundary conditions. The combination yields accurate results over a large range of particle-boundary separations, up to roughly one LB lattice spacing; as verified for the specific geometry of two parallel plates.
\item We estimate the raspberry to accurately reproduce hydrodynamics interactions for particle-particle and particle-surface separations greater than one LB lattice spacing. The separation is determined by the effective radii. Here, we consider the results of our rotational mobility experiments and translational mobility experiments in confinement significant, as there are no back-torques or back-forces applied to the fluid.~\cite{fischer15}
\end{itemize}
The latter estimate is not unreasonable, since the near-field effects of discretizing the surface into a finite number of MD beads become noticeable at comparable separations. There are, however, two caveats to the last result. (i) We have not verified that we obtain the correct behavior in a system with many particles that can undergo collision events. Here, we only extrapolate our particle-image results~\cite{fischer15} to particle-particle interactions. This point is left for future study. (ii) In the case of a rotating raspberry, for which we observed that the particle-image interactions are correct up to one lattice spacing, the motion and structure of the raspberry are such that the fluid is constantly exposed to different fluid-coupling points (the MD beads). This may remove artifacts caused by the discretization of the surface (or volume) that are more noticeable in the translational experiment, where the orientation is fixed.

From the above and our examination in Part I~\cite{fischer15}, it becomes clear that our `filling + fitting' procedure for the raspberry model is an excellent way to approximate fluid-particle coupling in a Stokes' liquid, using an LB algorithm.

\section*{\label{sec:ack}Acknowledgements}

J.d.G. acknowledges financial support by a ``Nederlandse Organisatie voor Wetenschappelijk Onderzoek'' (NWO) Rubicon Grant (\#680501210). We thank the ``Deutsche Forschungsgemeinschaft'' (DFG) for financial funding through the SPP 1726 ``Microswimmers -- from single particle motion to collective behavior''. We are also grateful to L. Helden, O. Hickey, and U. Schiller for useful discussions.

\end{document}